\documentclass[preprint]{revtex4}
\usepackage{amsfonts}
\usepackage{amssymb}
\usepackage{latexsym}
\usepackage{graphicx}
\usepackage{amsmath}
\usepackage{epsfig,subfig}
\usepackage{epstopdf}
\usepackage[active]{srcltx}

\begin{document}

\title{Macroscopic quantum behaviour of periodic quantum systems}
\author{A. Mart\'{\i}n-Ruiz}
\email{alberto.martin@nucleares.unam.mx}
\affiliation{Instituto de Ciencias Nucleares, Universidad Nacional Aut\'{o}noma de M\'{e}%
xico, 04510 M\'{e}xico, D.F., Mexico}
\author{J. Bernal}
\affiliation{Universidad Ju\'{a}rez Aut\'{o}noma de Tabasco. Divisi\'{o}n Acad\'{e}mica
de Ciencias B\'{a}sicas. Cunduac\'{a}n, Tabasco, M\'{e}xico.}
\author{Adri\'{a}n Carbajal-Dom\'{\i}nguez}
\affiliation{Universidad Ju\'{a}rez Aut\'{o}noma de Tabasco. Divisi\'{o}n Acad\'{e}mica
de Ciencias B\'{a}sicas. Cunduac\'{a}n, Tabasco, M\'{e}xico.}

\begin{abstract}
In this paper we introduce a simple procedure for computing the macroscopic
quantum behaviour of periodic quantum systems in the high energy regime. The
macroscopic quantum coherence is ascribed to a one-particle state, not to a
condensate of a many-particle system; and we are referring to a system of
high energy but with few degrees of freedom. We show that, in the first
order of approximation, the quantum probability distributions converge to
its classical counterparts in a clear fashion, and that the interference
effects are strongly suppressed. The harmonic oscillator provides a testing
ground for these ideas and yields excellent results.
\end{abstract}

\maketitle


\section{Introduction}

The superposition principle lies at the heart of quantum mechanics, and it
is one of its features that most distinctly marks the departure from
classical concepts \cite{Schlosshauer}. Formally, the superposition
principle is rooted in the linearity of the Hilbert space. A striking and
inevitable counterintuitive consequence of this principle is the phenomenon
of quantum interference. The simplest kind of interference is displayed in
the famous double-slit experiments, first performed by Young in 1802 with
light, then duplicated during the last century with all sorts of material
particles.

At the macroscopic scale all happens as if, by a conspiracy of nature, the
naked quantum is hidden, leaving us with an apparent world described
consistently in a classical language \cite{Haroche}. The question that
naturally arises is if quantum effects can be observed at macroscopic level.
A classical illustration of the conflict between the existence of quantum
superpositions and our real-world experience (of observation and
measurement) is the "Schr\"{o}dinger's cat", in which a cat is put in a
quantum superposition of alive and dead states. A widely accepted
explanation nowadays for the appearance of classical like features from an
underlying quantum world is the environment induced decoherence approach 
\cite{Joos, Zurek1, Zurek2, Giulini}. According to this theory, coupling to
a large number of degrees of freedom (the environment) results in a loss of
quantum coherence which leads to emergent classicality.

In this paper we introduce a simple procedure to compute the high energy
regime of a general density matrix for periodic quantum systems. We show
that, in the first order of approximation, the quantum probability
distribution converges to its classical counterpart in a clear fashion, and
that the interference effects are strongly suppressed. The harmonic
oscillator provides a testing ground for these ideas as we will illustrate.
We demonstrates that the classical features emerges from its quantum
description in the high energy regime. This problem has been considered
before by Cabrera and Kiwi \cite{Cabrera}. In this work, they use purely
quantum-mechanical results to analyze (by inspection) the amplitude of the
oscillations and the spatial autocorrelation function for large quantum
numbers. They conclude that even for arbitrarily high quantum numbers, a
superposition of (a few) eigenstates retains quantum effects. The
originality of our paper lies in the fact that we propose a well-founded
mathematical procedure for calculating the macroscopic manifestations of
quantum behaviour.

The remainder of the paper is organized as follows. In Sec. II, we introduce
the general procedure. The results for the harmonic oscillator are presented
in Section III. Finally, some conclusions and remarks are given in Section
IV.

\section{General procedure}

It is generally accepted that the classical and quantum probability density
functions for periodic systems approach each other in a locally averaged
sense when the principal quantum number becomes large \cite{Liboff,
Auerbach, Yoder, Robinett}, i.e. 
\begin{equation}
\rho ^{CL}\left( x\right) =\lim_{n\gg 1}\frac{1}{2\epsilon _{n}}%
\int_{x-\epsilon _{n}}^{x+\epsilon _{n}}\rho _{n}^{QM}\left( y\right) dy\,%
\text{,}  \label{L.AVERAGE}
\end{equation}%
where the interval $\epsilon _{n}$ decreases with increasing the quantum
number $n$. For a brief discussion of the meaning of the classical
probability distribution see Ref.\cite{Robinett}. The local averaging of the
infinite square well potential is reported in \cite{Auerbach} and \cite%
{Robinett}. In most cases, Eq.(\ref{L.AVERAGE}) is very difficult to
calculate analitically. We briefly review an alternative procedure to
compute the local averages appearing Eq.(\ref{L.AVERAGE}) \cite{Bernal,
Martin}. Supported by the harmonic analysis criteria, the authors write the
classical and quantum distributions as a Fourier expansion,%
\begin{eqnarray}
\rho _{n}^{QM}\left( x\right) &=&\int f_{n}^{QM}\left( p\right) e^{i\frac{px%
}{\hbar }}dp,  \label{FOURIER} \\
\rho ^{CL}\left( x\right) &=&\int f^{CL}\left( p\right) e^{i\frac{px}{\hbar }%
}dp,  \notag
\end{eqnarray}%
where $f_{n}^{QM}\left( p\right) $ and $f^{CL}\left( p\right) $ are the
Fourier coefficients of each expansion respectively. An immediate
consequence of Eqs. (\ref{L.AVERAGE}) and (\ref{FOURIER}) is that the
Fourier coefficients have a similar behaviour for $n$ large,%
\begin{equation}
f_{n}^{QM}\left( p\right) \sim f^{CL}\left( p\right) .  \label{ASYMP}
\end{equation}%
Note that Planck's constant keep a finite value, so $\hbar $-dependent
corrections may arise in Eq.(\ref{ASYMP}). This implies that even at
macroscopic level the Heisenberg's theorem works. Finally calculating the
inverse Fourier transform of the asymptotic Fourier coefficients we obtain,\
at least in a first approximation, the classical probability density.
Analytical results for the simplest quantum systems were reported \cite%
{Bernal, Martin}.

Now we focus on the general problem. Let us consider a physical system
described completely by the Hamiltonian $\hat{H}$. Let $\left\{ \left\vert
\psi _{n}\right\rangle \right\} $ be an orthonormal basis of eigenvectors of 
$\hat{H}$ with eigenvalues $E_{n}$. If the initial state $\left\vert \Psi
\left( 0\right) \right\rangle $ has coefficients $c_{n}=\left\langle \psi
_{n}|\Psi \left( 0\right) \right\rangle $, then the state at a later time $t$%
, according to Schr\"{o}dinger equation, is given by%
\begin{equation}
\left\vert \Psi \left( t\right) \right\rangle =\sum_{n}c_{n}e^{-i\frac{E_{n}t%
}{\hbar }}\left\vert \psi _{n}\right\rangle .  \label{STATE}
\end{equation}%
The corresponding density matrix reads%
\begin{equation}
\hat{\rho}\left( t\right) =\left\vert \Psi \left( t\right) \right\rangle
\left\langle \Psi \left( t\right) \right\vert
=\sum\limits_{n,m}c_{n}c_{m}^{\ast }e^{-i\frac{\left( E_{n}-E_{m}\right) t}{%
\hbar }}\left\vert \psi _{n}\right\rangle \left\langle \psi _{m}\right\vert .
\label{DENS MAT}
\end{equation}%
The terms $m\neq n$ embody the quantum coherence between the different
components $\left\vert \psi _{n}\right\rangle $. Accordingly, they are
usually referred to as interference terms, or off-diagonal terms.
Consequently, the expectation value of the arbitrary observable $\mathcal{O}$
is given by 
\begin{equation}
\left\langle \mathcal{O}\right\rangle =Tr\left[ \mathcal{O}\hat{\rho}\left(
t\right) \right] =\sum\limits_{n,m}c_{n}c_{m}^{\ast }\mathcal{O}_{m,n}e^{-i%
\frac{\left( E_{n}-E_{m}\right) t}{\hbar }},  \label{EXP VAL}
\end{equation}%
with $\mathcal{O}_{m,n}\equiv \left\langle \psi _{m}\right\vert \mathcal{O}%
\left\vert \psi _{n}\right\rangle $.

To explain our procedure, let us consider the coordinate representation of
the matrix density. Its components are given by%
\begin{equation}
\rho _{n,m}\left( x,t\right) =c_{n}c_{m}^{\ast }\tilde{\rho}_{n,m}\left(
x\right) e^{-i\frac{\left( E_{n}-E_{m}\right) t}{\hbar }},
\label{COMP DENS MAT}
\end{equation}%
with $\tilde{\rho}_{n,m}\left( x\right) \equiv \psi _{n}\left( x\right) \psi
_{m}^{\ast }\left( x\right) $. It is well known that space and time play
fundamentally different roles in quantum mechanics: whereas position is
represented by a hermitian operator, time is represented by a $c$-number 
\cite{Neumann}. Then we study separately the roles of space and time in Eq.(%
\ref{COMP DENS MAT}). We first consider the spacial behaviour.

It is well known that for periodic quantum systems, nodes are always present
in the density matrix (by means of $\tilde{\rho}_{n,m}\left( x\right) $) for
arbitrarily large quantum numbers, and thus the study of its macroscopic
behaviour naturally implies an average process. To this end we extend the
Fourier expansion in Eq.(\ref{FOURIER}) to the spatial components of the
matrix density (\ref{COMP DENS MAT}),%
\begin{equation}
\tilde{\rho}_{n,m}\left( x\right) =\int f_{n,m}\left( p\right) e^{i\frac{px}{%
\hbar }}dp,  \label{FOURIER DENS MAT}
\end{equation}%
where $f_{n,m}\left( p\right) $ are the Fourier coefficients of the
expansion. Physically the Fourier coefficients $f_{n,m}\left( p\right) $ are
the convolution $\phi _{n}\left( p\right) \ast \phi _{m}\left( p\right) $,
where $\phi _{n}\left( p\right) $ is the wavefunction in momentum
representation. Subsequently we consider the spacial asymptotic behaviour of
the Fourier coefficients for large quantum numbers, and finally its inverse
Fourier transform gives the desired macroscopic behaviour of $\tilde{\rho}%
_{n,m}\left( x\right) $. According to our macroscopic-world experience, the
interference effects are completely suppressed. For instance, no one has
ever seen a ball going through two directions at once \cite{Haroche}.
Therefore, the asymptotics of $\tilde{\rho}_{n,m}\left( x\right) $, for $%
n\neq m$, should be strongly suppressed.

We now study the temporal behaviour. We observe that Eq.(\ref{COMP DENS MAT}%
) is an almost periodic function of the time (except for the diagonal
terms), even for arbitrarily high energies. It is clear that the
off-diagonal terms are a rapidly oscillating functions of the time when the
energies differ significantly, and a slowly varying functions if the
energies are close each other. This means that in the macroscopic regime
becomes important only the interference between states with high energies $%
E_{n}$ and $E_{m}$, such that $\left\vert E_{n}-E_{m}\right\vert
=E_{\upsilon }$ and $E_{\upsilon }\ll E_{n}$. Under these conditions the
quantum-mechanical frequencies goes to its classical counterpart \cite%
{Lindenbaum}, i.e.,%
\begin{equation}
\omega _{\upsilon }^{QM}\equiv \frac{E_{n+\upsilon }-E_{n}}{\hbar }\approx
2\pi \upsilon \frac{\partial E_{n}}{\partial J}=\upsilon \omega ^{CL}\text{,}
\label{frequency}
\end{equation}%
where we used the standard definition $\omega ^{CL}\equiv 2\pi \frac{%
\partial E_{n}}{\partial J}$, where $J=nh$ is the action \cite{Goldstein}.

We conclude this section with the final expression for the \textit{%
macroscopic density matrix} in coordinate representation, i.e.,%
\begin{equation}
\rho \left( x,t\right) \sim \sum\limits_{n,\upsilon }c_{n}c_{n-\upsilon
}^{\ast }\tilde{\rho}_{n,n-\upsilon }\left( x\right) e^{-i\upsilon \omega
^{CL}t}\text{.}  \label{MAC DENS MAT}
\end{equation}%
where we have included the spatial and temporal analysis described above. We
observe that (\ref{MAC DENS MAT}) is the Fourier expansion of a classical
function, which can be inmediately identified with the classical probability
distribution $\rho ^{CL}\left[ x\left( t\right) \right] $. Note that the
long-time behavior of $\rho \left( x,t\right) $ converges towards $%
\sum\limits_{n}\left\vert c_{n}\right\vert ^{2}\tilde{\rho}_{n,n}\left(
x\right) $, in agreement with the classical result. We point out that in our
work the macroscopic quantum coherence is ascribed to a one-particle state,
not to a condensate of a many-particle system. We are referring to a system
of high energy, but with few degrees of freedom. In the next section we
present analytical results for the harmonic oscillator.

\section{Macroscopic density matrix of the harmonic oscillator}

The harmonic oscillator provides a testing ground for these ideas as we now
illustrate. We first consider again the spacial behaviour. The energy
eigenfunctions and eigenvalues of the corresponding Schr\"{o}dinger equation
are well known. The quantized energies are given by%
\begin{equation}
E_{n}=\hbar \omega \left( n+\frac{1}{2}\right)  \label{ENERGY}
\end{equation}%
with $n\in 
\mathbb{Z}
^{+}$, while the stationary states can be written as follows%
\begin{equation}
\psi _{n}\left( x\right) =\sqrt[4]{\frac{\alpha }{\pi }}\frac{1}{\sqrt{%
2^{n}n!}}H_{n}\left( \sqrt{\alpha }x\right) e^{-\frac{\alpha x^{2}}{2}}
\label{STATE HO}
\end{equation}%
where $\alpha =\frac{m\omega }{\hbar }$. The spatial components of the
matrix density are given by

\begin{equation}
\tilde{\rho}_{n,m}\left( x\right) =\sqrt{\frac{\alpha }{\pi }}\frac{1}{\sqrt{%
2^{n}n!2^{m}m!}}H_{n}\left( \sqrt{\alpha }x\right) H_{m}\left( \sqrt{\alpha }%
x\right) e^{-\alpha x^{2}}\text{.}  \label{OFF DIAG HO}
\end{equation}%
Figure 1 illustrates the spatial components $\tilde{\rho}_{n,m}\left(
x\right) $ for different values of $n$ and $m$. In Figs. 1(a), 1(b) and 1(c)
we first consider the smooth oscillatory quantum regime for $n=1$ and $%
m=1,10,~20$, respectively. When increasing the quantum numbers, $\tilde{\rho}%
_{n,m}\left( x\right) $ becomes a rapidly oscillatory functions around the $%
x $-axis. This is illustrated in Figs. 1(d), 1(e) and 1(f) for $n=100$ and $%
m=100,125,150$, respectively. As we anticipated, nodes are always present
for arbitrarily large quantum numbers. Note that the number of nodes
increase with decreasing the difference $n-m$. It is clear that, after the
local averaging process, the off-diagonal components will be strongly
suppresed compared with the diagonal terms. Now we focus on applying the
procedure presented in Sec.I.

\begin{figure}[tbp]
\vspace{0.5 cm}
\par
\begin{center}
\subfloat[]{\includegraphics[scale=0.8, natwidth=640,
natheight=480]{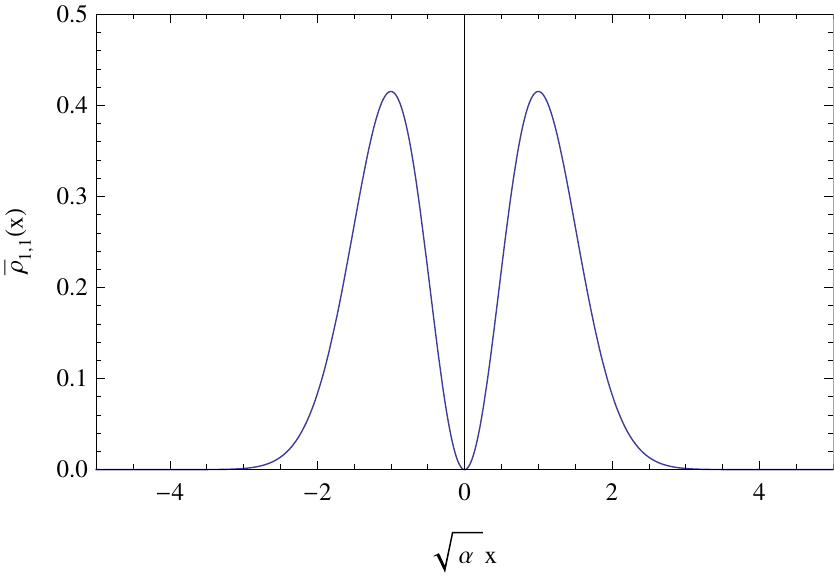}}%
\subfloat[]{\includegraphics[scale=0.8,
natwidth=640, natheight=480]{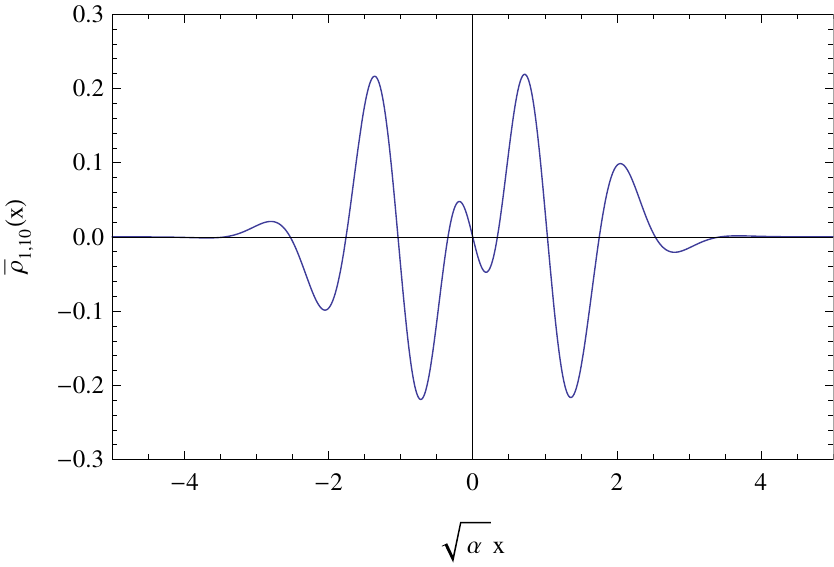}}\\[0pt]
\subfloat[]{\includegraphics[scale=0.8, natwidth=640,
natheight=480]{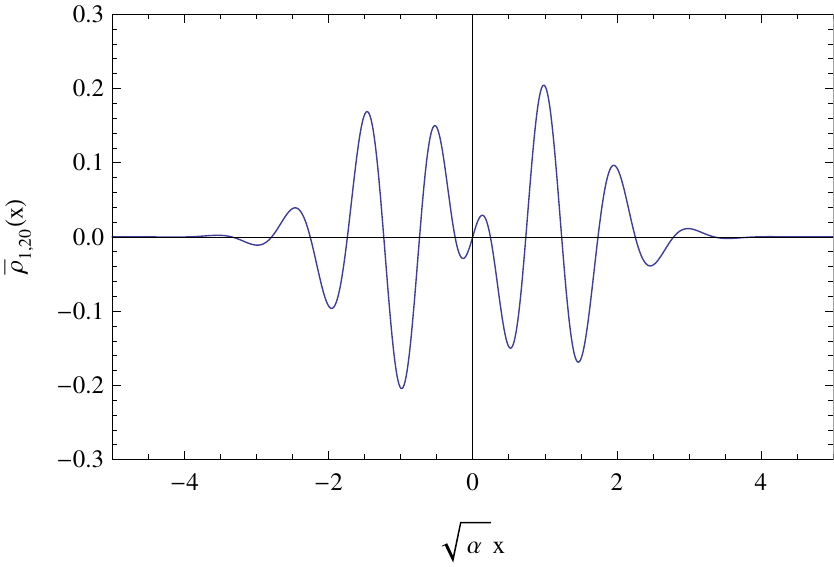}}%
\subfloat[]{\includegraphics[scale=0.8,
natwidth=640, natheight=480]{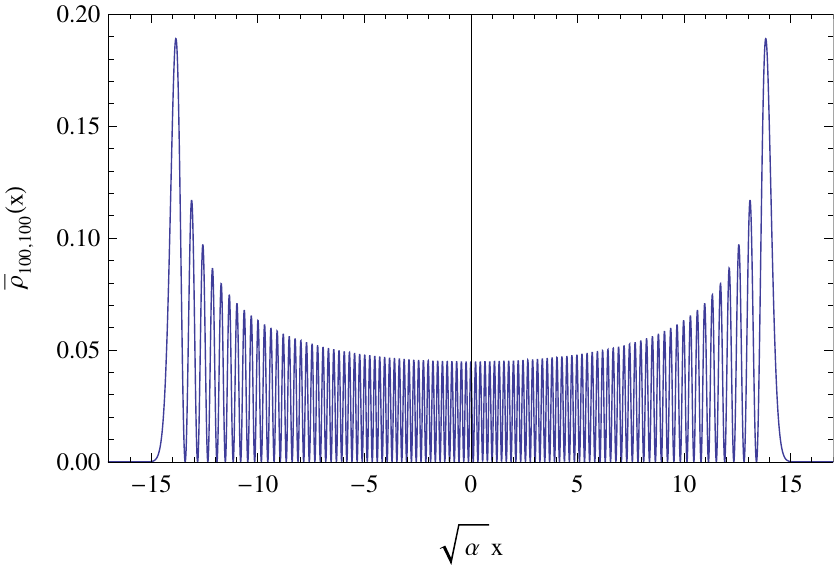}}\\[0pt]
\subfloat[]{\includegraphics[scale=0.8, natwidth=640,
natheight=480]{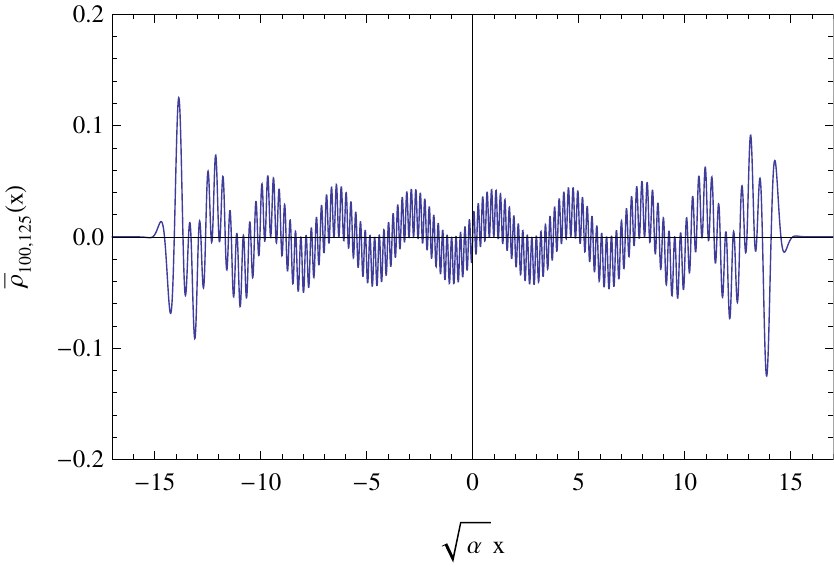}}%
\subfloat[]{\includegraphics[scale=0.8,
natwidth=640, natheight=480]{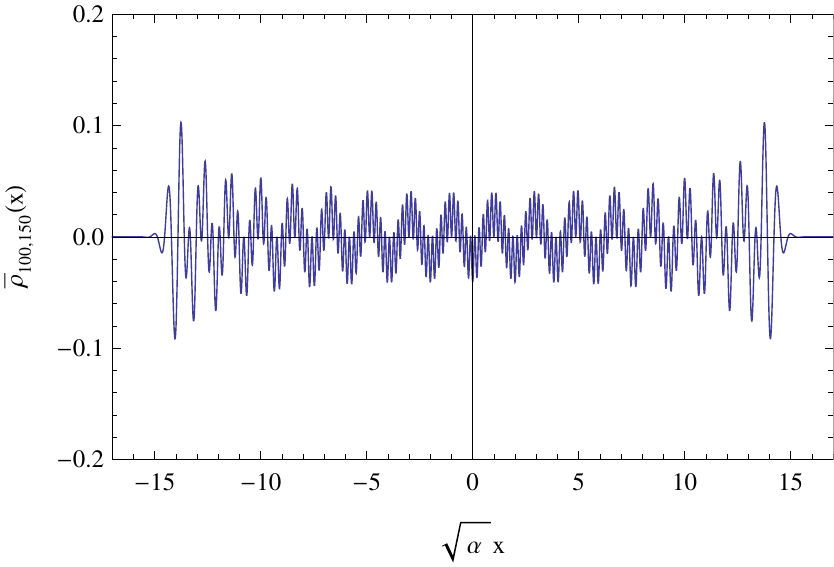}}
\end{center}
\caption{Spatial behaviour of $\bar{\protect\rho}_{n,m}\left( x\right)
\equiv \frac{\tilde{\protect\rho}_{n,m}\left( x\right) }{\protect\sqrt{%
\protect\alpha }}$. \ Graphs (a), (b) and (c) corresponds to $n=1$ and $%
m=1,10,~20$, respectively. The high energy regime is illustrated in graphs
(d), (e) and (f) for $n=100$ and $m=100,125,150$, respectively.}
\end{figure}

We first calculate the Fourier coefficients appearing in Eq.(\ref{FOURIER
DENS MAT}). The corresponding inverse Fourier transform is reported in many
handbooks of mathematical functions \cite{Abramowitz, Gradshteyn}: 
\begin{equation}
f_{n,m}\left( p\right) =\left( -i\right) ^{n-m}\sqrt{\frac{m!}{n!}}e^{-\frac{%
\xi _{0}^{2}}{2}}\xi _{0}^{n-m}L_{m}^{n-m}\left( \xi _{0}^{2}\right)
\label{FOURIER2 HO}
\end{equation}%
where $n\geq m$, $\xi _{0}\equiv \frac{p}{\sqrt{2m\omega \hbar }}$ and $%
L_{m}^{n-m}$ is an associated Laguerre polynomial. In this expression it can
be seen that the correlation between two wave functions increases as the
difference $n-m$ decrease.

The asymptotic behaviour of $f_{n,m}\left( p\right) $, for $n$ and $m$
large, is also well known. Szeg\"{o} finds the following iterative relation
for $n$ large \cite{Szego}:%
\begin{eqnarray}
u\left( x\right) &=&e^{-\frac{x^{2}}{2}}x^{\upsilon }L_{n}^{\upsilon }\left(
x^{2}\right) \sim \frac{\Gamma \left( n+\upsilon +1\right) }{N^{\frac{%
\upsilon }{2}}n!}J_{\upsilon }\left( 2\sqrt{N}x\right) -\frac{\pi }{2}\times
\label{SZEGO} \\
&&\int_{0}^{x}y^{3}u\left( y\right) \left[ J_{\upsilon }\left( 2\sqrt{N}%
x\right) Y_{\upsilon }\left( 2\sqrt{N}y\right) -Y_{\upsilon }\left( 2\sqrt{N}%
x\right) J_{\upsilon }\left( 2\sqrt{N}y\right) \right] dy  \notag
\end{eqnarray}%
where $J_{\upsilon }$ and $Y_{\upsilon }$ are the usual Bessel functions of
the first and second kind respectively, and $N=n+\frac{\upsilon +1}{2}$. Szeg%
\"{o} also shows that the iteration terms are strongly suppressed compared
with the first order of approximation in the limit $N\rightarrow \infty $.
In this paper we will consider only the first order of approximation,
however the higher orders of approximation follows immediately from Eq.(\ref%
{SZEGO}). The asymptotic behaviour of the Fourier coefficients is then%
\begin{equation}
f_{n,n-\upsilon }\left( p\right) \sim \frac{\left( -i\right) ^{\upsilon }}{%
\left( 1-\frac{\upsilon -1}{2n}\right) ^{^{\frac{\upsilon }{2}}}}J_{\upsilon
}\left( 2\sqrt{N}\xi _{0}\right)  \label{ASYMP FOURIER HO}
\end{equation}%
with $N=n-\frac{\upsilon -1}{2}$ and $\upsilon =n-m$, such that $n\gg
\upsilon \geq 0$. In Eq.(\ref{ASYMP FOURIER HO}) we have used the
approximation $\frac{n!}{m!}\sim n^{\upsilon }$. From Eq.(\ref{ASYMP FOURIER
HO}) it follows that, in the asymptotic Fourier coefficients matrix, the
order of the Bessel function increases as we move away from the main
diagonal. For example, in the main diagonal we have $J_{0}\left( 2\sqrt{n+%
\frac{1}{2}}\xi _{0}\right) $, in the secondary diagonals we have firstly $%
-iJ_{1}\left( 2\sqrt{n}\xi _{0}\right) $, secondly $-\left( 1-\frac{1}{2n}%
\right) ^{-\frac{1}{^{^{2}}}}J_{2}\left( 2\sqrt{n+\frac{1}{2}}\xi
_{0}\right) $, and so on. Physically this means that in the high energy
regime the interference between two states becomes important when its
energies are close each other, as anticipated.

Finally the macroscopic behaviour of $\tilde{\rho}_{n,n-\upsilon }\left(
x\right) $ is obtained by substituting the Eq.(\ref{ASYMP FOURIER HO}) into
Eq.(\ref{FOURIER DENS MAT}) and calculating the resulting Fourier transform.
In Refs.\cite{Erdelyi} and \cite{Kammler} is reported the Fourier
transformation of Bessel functions. Our final expression for the asymptotic
behaviour of $\tilde{\rho}_{n,n-\upsilon }\left( x\right) $ for large
quantum numbers is then%
\begin{equation}
\tilde{\rho}_{n,n-\upsilon }\left( x\right) \sim \frac{1}{\left( 1-\frac{%
\upsilon -1}{2n}\right) ^{^{\frac{\upsilon }{2}}}}\frac{T_{\upsilon }\left( 
\frac{x}{\chi _{n,n-\upsilon }}\right) }{\pi \sqrt{\chi _{n,n-\upsilon
}^{2}-x^{2}}}\mathrm{Rect}\left( \frac{x}{2\chi _{n,n-\upsilon }}\right)
\label{ASYMP OFF DIAG HO}
\end{equation}%
where\ $\chi _{n,n-\upsilon }\equiv \sqrt{\frac{2N\hbar }{m\omega }}$, $%
T_{\upsilon }$ is the Chebyshev polynomial of the first kind and $\mathrm{{%
\mathrm{Rect}}}$ is the rectangular function. Note that the rectangular
function restricts the domain of (\ref{ASYMP OFF DIAG HO}) to $x\in \left[
-\chi _{n,n-\upsilon },\chi _{n,n-\upsilon }\right] $.

We observe that the diagonal terms in (\ref{ASYMP OFF DIAG HO}) are simply%
\begin{equation}
\tilde{\rho}_{n,n}\left( x\right) \sim \frac{1}{\pi \sqrt{x_{n}^{2}-x^{2}}}%
\mathrm{{\mathrm{Rect}}}\left( \frac{x}{2x_{0}}\right) \text{,}
\label{ASYMP DIAG HO}
\end{equation}%
with $x_{n}\equiv \chi _{n,n}=\sqrt{\frac{2\left( n+\frac{1}{2}\right) \hbar 
}{m\omega }}$. Note that $\tilde{\rho}_{n,n}\left( x\right) $ coincides with
the probability density of a classical harmonic oscillator with amplitude $%
x_{n}$. Rearranging the expression for $\chi _{n,n}$ we obtain $\frac{1}{2}%
m\omega ^{2}x_{n}^{2}=\hbar \omega \left( n+\frac{1}{2}\right) $, which is
the well known expression for the classical energy. Therefore we have shown
that the classical features (e.g. amplitude, energy, probability
distribution and the confinement effect) naturally emerge as the first order
of approximation of quantum mechanics in the high energy regime. When
considering higher orders of approximation in (\ref{ASYMP DIAG HO}) we
observe macroscopic quantum behaviour. For example, a residual oscillatory
behaviour (in all space) is retained in $\tilde{\rho}_{n,n}\left( x\right) $
even for arbitrarily high quantum number $n$ \cite{Bernal}. This is because $%
\hbar $ keeps a finite value, and thus the Heisenberg's theorem still works.
The exact classical result is recovered if we take $\hbar \rightarrow 0$,
however $\hbar $ is a fundamental constant of nature whose numerical value
although small is not zero.

The exact classical limit requires the off-diagonal terms equal to zero
(means no interference), however according to our result (\ref{ASYMP OFF
DIAG HO}) this is never attained. In figures 2 we present the asymptotic
behaviour of $\tilde{\rho}_{n,n-\upsilon }\left( x\right) $ for $n=10000$
and different values of $\upsilon $. Figs. 2(a) and 2(b) depict the
interference between\ neighboring\textit{\ }states with $\upsilon =1,2$
respectively, while in figs. 2(c) and 2(d) we consider states with $\upsilon
=50,100$ respectively.\ Note that in all cases, although the interference
effect is very small (less than $0.01$) is not zero. Also it can be seen
that the nodes increase with increasing $\upsilon $, so that the expectation
values $\left\langle \psi _{m}\right\vert \mathcal{O}\left\vert \psi
_{n}\right\rangle $ of a smooth function\ $\mathcal{O}\left( x\right) $
converge more rapidly to zero for $\upsilon $ large. This means that at high
energies the interference becomes important only for neighboring\textit{\ }%
states.

\begin{figure}[tbp]
\vspace{0.5 cm}
\par
\begin{center}
\subfloat[]{\includegraphics[scale=0.8, natwidth=640,
natheight=480]{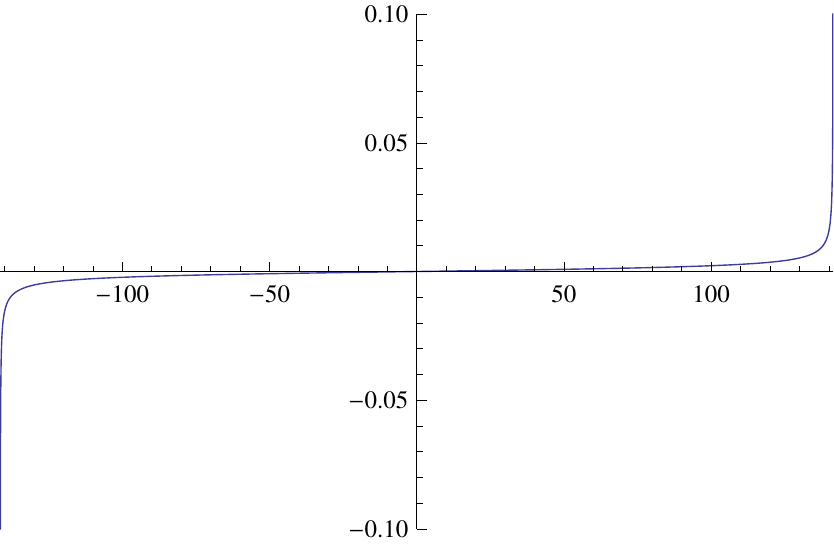}}\qquad 
\subfloat[]{\includegraphics[scale=0.8,
natwidth=640, natheight=480]{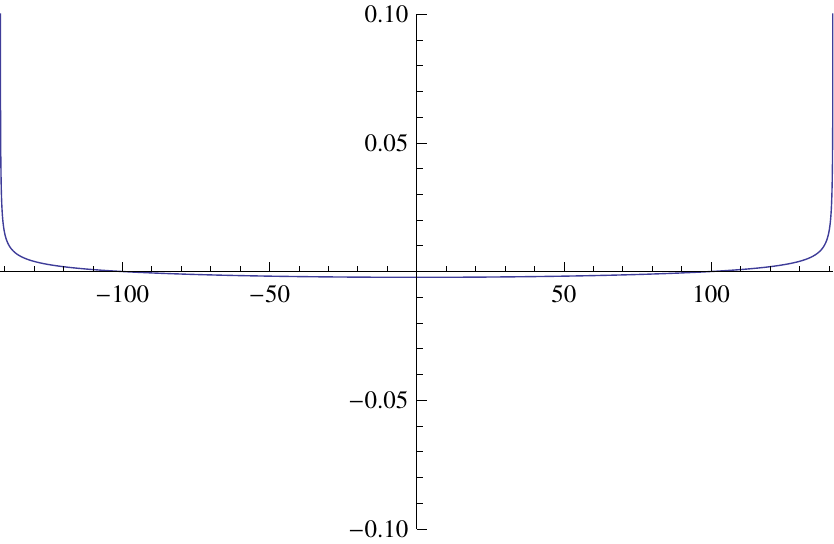}}\\[0pt]
\subfloat[]{\includegraphics[scale=0.8, natwidth=640,
natheight=480]{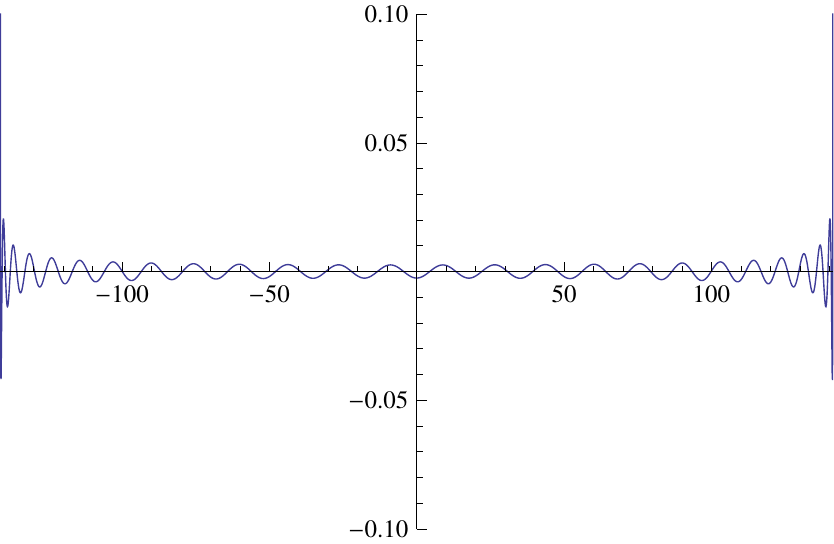}}\qquad 
\subfloat[]{\includegraphics[scale=0.8,
natwidth=640, natheight=480]{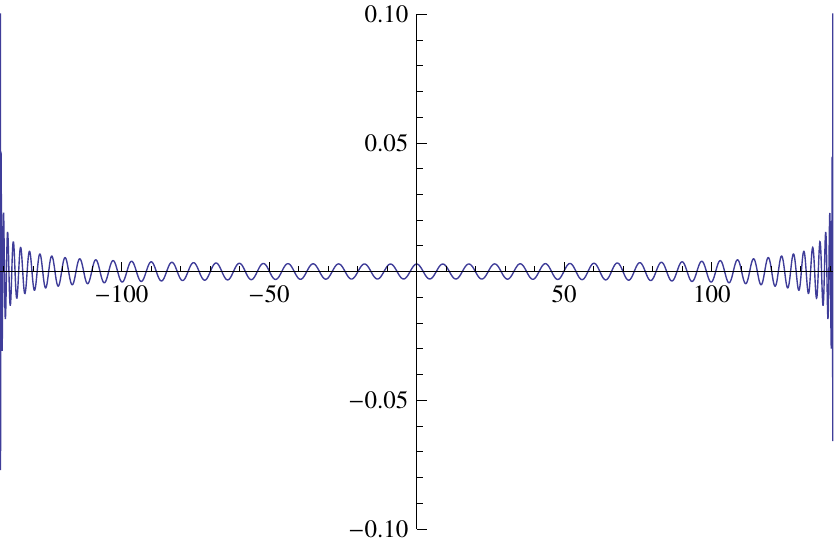}}
\end{center}
\caption{Asymptotic spatial behaviour of $\bar{\protect\rho}_{n,m}\left(
x\right) \equiv \frac{\tilde{\protect\rho}_{n,m}\left( x\right) }{\protect%
\sqrt{\protect\alpha }}$ for $n=10000$. Graphs (a) and (b) depict the
interference between\ neighboring\textit{\ }states with $\protect\upsilon %
=1,2,$ respectively, while in graphs (c) and (d) we consider states with $%
\protect\upsilon =50,100,$ respectively.}
\end{figure}

Regarding to the temporal behaviour, from Eq.(\ref{COMP DENS MAT}) it
follows that the off-diagonal temporal terms in the matrix density are
simply $e^{-i\upsilon \omega t}$, were we used (\ref{ENERGY}). We observe
that the frequency of the temporal oscillations increase as we move away
from the main diagonal. In this case the high energy temporal behaviour is
exactly the same as its low energy counterpart (with the same $\upsilon $).
In the macroscopic regime $\upsilon \ll n$ therefore $e^{-i\upsilon \omega
t} $ is a slowly varying function.

To complete this section we write an asymptotic expression for the
expectation value of an arbitrary observable $\mathcal{O}$. According to our
results $\left\langle \mathcal{O}\right\rangle $ will be, at first order of
approximation, the classical expectation value $\left\langle \mathcal{O}%
\right\rangle ^{CL}$ plus corrections coming from the possible interferece
between states, i.e.

\begin{equation}
\left\langle \mathcal{O}\right\rangle \sim \left\langle \mathcal{O}%
\right\rangle ^{CL}+\sum\limits_{n,\upsilon }\frac{c_{n}c_{n-\upsilon
}^{\ast }e^{-i\upsilon \omega t}}{\left( 1-\frac{\upsilon -1}{2n}\right) ^{^{%
\frac{\upsilon }{2}}}}\int_{-1}^{+1}\mathcal{O}\left( \chi _{n,n-\upsilon
}\xi \right) \frac{T_{\upsilon }\left( \xi \right) }{\pi \sqrt{1-\xi ^{2}}}%
d\xi ,  \label{ASYMP EXP VAL}
\end{equation}%
where $\left\langle \mathcal{O}\right\rangle ^{CL}=\sum\limits_{n}\left\vert
c_{n}\right\vert ^{2}\int \mathcal{O}\left( x\right) \tilde{\rho}%
_{n,n}\left( x\right) .$ We can also evaluate higher orders of approximation
in a simple fashion \cite{Szego}, however these are strongly suppressed
compared with Eq.(\ref{ASYMP EXP VAL}). In practice only a few number of
terms are important in the summation. Because the properties of Chebyshev
polynomials, if $\mathcal{O}\left( x\right) $ is a polynomial function of
order $k$, then the integral in Eq.(\ref{ASYMP EXP VAL}) vanishes for $%
k>\upsilon $ \cite{Abramowitz}. Therefore $\left\langle x\right\rangle $ and 
$\left\langle x^{2}\right\rangle $ requires only $\upsilon =1,2$,
respectively. We point out that the ergodic behaviour of Eq.(\ref{ASYMP EXP
VAL}) yields the correct classical expectation value.

\section{Final remarks}

It is well known that the concepts of classical and macroscopic systems are
distinct, as the existence of macroscopic quantum phenomena (such as
superconductivity) demonstrates, but the behaviour of most macroscopic
systems can be described by classical theories \cite{Ballentine}. In this
paper we have shown that quantum mechanics is applicable in every scale of
nature, and the macroscopic regime emerge as a consequence of its high
energy behaviour. Quantum effects remains at this level (called macroscopic
quantum behaviour), as the interference between quantum states. It would be
interesting to test these effects with real quantum systems approaching the
microscopic-macroscopic boundary, as Rydberg atoms or neutron interferometry
for example. At higher energies these macroscopic quantum effects are so
strongly suppressed that it is impossible to detect, leaving us with an
apparent world described consistently in a classical language. With the
appropriate experimental devices such effects should be observed even in our
real-world experience, however nowadays it is impossible.

Technical difficulties in the calculation of the Kepler problem are greater
than in the simple case which we have treated here, however we can definetly
foresee that our procedure gives its correct macroscopic behaviour. Even
though our approach gives the correct classical results for periodic quantum
systems, it is far from the general solution to the classical limit problem.
Several other questions remain to be resolved as the study of unbound
systems and entanglement, but is not clear in the framework adopted here.
The environment induced decoherence approach successfully resolves these
problems.

Based in our results, in our future research we intend to concentrate on the
definition of the classical regime, which is of considerable importance in
the study of quantum chaos.

B. Buck and C. V. Sukumar, Phys. Lett. 83A, 132 (1981).

American Journal of Physics -- May 2006 -- Volume 74, Issue 5, pp. 404

\end{document}